\documentstyle[aps,prl,floats,epsfig]{revtex}	

\title{OBSERVATION OF SHELL STRUCTURE IN SODIUM NANOWIRES}
\author{A.I. Yanson$^{\star}$, I.K. Yanson$^{\star, \dagger}$ \& J.M. van Ruitenbeek$^{\star}$}
\address{$^{\star}$Kamerlingh Onnes Laboratorium,\\ Universiteit Leiden,
PO Box 9504, NL-2300 RA Leiden, The Netherlands\\
$^{\dagger}$B. Verkin Institute for Low Temperature Physics and Engineering,\\ National Academy of Sciences, 310164, Kharkiv, Ukraine} 

\begin{document}
\draft

\twocolumn[\hsize\textwidth\columnwidth\hsize\csname@twocolumnfalse\endcsname

\maketitle

\begin{abstract}
{\bf The quantum states of a system of particles in a finite spatial domain 
in general consist of a set of discrete energy eigenvalues; these are usually grouped into bunches of degenerate or close-lying levels\cite{Balian}, called shells. In fermionic systems, this gives rise to a local minimum in the total energy when all the states of a given shell are occupied. In particular, the closed-shell electronic configuration of the noble gases produces their exceptional stability. Shell effects have previously been observed for protons and neutrons in nuclei and for clusters of metal atoms \cite{Bohr,de Heer,Brack93}. Here we report the observation of shell effects in an open system -- a sodium metal nanowire connecting two bulk sodium metal electrodes, which are progressively pulled apart. We measure oscillations in the
statistical distribution of conductance values, for contact cross-sections 
containing up to a hundred atoms or more. The period follows the law expected
for the electronic shell-closure effects, similar to the abundance peaks at `magic numbers' of atoms in metal clusters \cite{de Heer,Brack93}. }
\end{abstract}


\vskip2pc]

\narrowtext

\newcommand{\av}[1]{\mbox{$\langle #1 \rangle$}}

Metallic constrictions, in the form of nanowires connecting two bulk
metal electrodes, have recently been studied down to sizes of a single atom
in cross section by means of scanning tunnelling microscopy (STM) and mechanically controllable break-junctions (MCB)\cite{Agrait93,Muller92}. By indenting one electrode into another and then separating them, a stepwise decrease in electrical conductance is observed, down to the breakpoint when arriving at the last atom. Each scan of the dependence of conductance $G$ on the elongation $d$ is individual in detail, as the atomic configuration of each contact may be widely different. However, statistically, many scans together produce a histogram of the probability for observing a given conductance value,
which is quite reproducible for a given metal and for fixed experimental
parameters.

A simple description of the electronic properties of metallic nanowires is
expected to work best for monovalent free-electron-like metals. The alkali
metals are most suitable, as the bulk electronic states are very well
described by free particles in an isotropic homogeneous positive background.
In a previous experiment on sodium point contacts \cite{Krans95} a histogram
was obtained, which showed pronounced peaks near 1, 3, 5 and 6 times the
quantum unit of conductance, $G_{0}=2e^{2}/h$ (where $e$ is the charge on an electron and $h$ is Planck's constant). This is exactly the series of
quantum numbers expected for the conductance through an ideal cylindrical
conductor. We have now extended these experiments to higher temperatures and a much wider range of conductance values. 

The MCB technique \cite{Muller92} was adapted for the study of nanowires of
the highly reactive alkali metals following Ref. \cite{Krans95} (see Fig.~1).
Scans were taken continuously by ramping the displacement $d$ of the
electrodes with respect to each other, using the piezo-electric driver. Each
individual curve of conductance versus displacement, $G(d)$, was recorded in
$\sim$0.1~seconds from the highest conductance into the tunnelling
regime. Histograms of conductance values were accumulated automatically involving $10^{3}$--$10^{5}$ individual scans, and for a number of
sample temperatures between 4.2~K and 100~K. Reproducibility and
reversibility were verified by periodically returning to the same measuring
conditions. Here we present results for sodium, but similar results were obtained for lithium and potassium.

\begin{figure}[!b]
\begin{center}
\leavevmode
\epsfig{figure=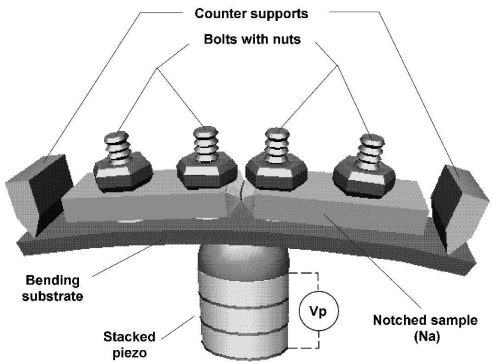,width=6.5cm}
\end{center}
\caption{Schematic view of the MCB technique for alkali metals \protect\cite{Krans95}. While immersed in paraffin oil, the sample is pressed onto four 1~mm diameter brass bolts, which are glued onto the isolated substrate. Current and voltage leads are fixed with nuts onto the bolts. A notch is cut into the sample at the centre. This assembly is mounted inside a vacuum can, which is immersed in liquid helium. By bending the substrate at 4.2\,K in vacuum, the sample is broken at the notch. Two fresh surfaces are exposed, and their distance is adjusted by changing the force on the substrate, employing a piezo element for fine control. The temperature of the sample is controlled by a heater and thermometer, which are fixed at the bottom of the substrate. The cryo-pumping action of the low-temperature environment ensures that the surfaces are not polluted by adsorbates. The purity of the sodium metal was at least 99.9\%. The conductance was recorded using a standard d.c. voltage bias 
four-terminal technique, measuring the current with 16-bit resolution. The drift and calibration of the current-to-voltage converter was verified against standard resistors corresponding to 1, 10 and 100~$G_0$, leading to an overall accuracy in the conductance better than 1\% for $G>10G_0$. }
\label{fig:fig1} 
\end{figure}

Figure~2 shows the temperature dependence of the histograms for sodium in the
range from 0 to 20~$G_{0}$. In order to compare different graphs, the amplitude has been normalised by the area under each graph. The histograms at low temperatures and low conductances are similar to those given in ref.~\onlinecite{Krans95}. We clearly recognise the familiar sharp peaks below 6~$G_{0}$, while at higher conductances a number of rather wide maxima are found. With increasing temperature we observe a gradual decrease in amplitude of the lower conductance peaks, while the high-conductance peaks dramatically sharpen and increase in amplitude. 

\begin{figure}[!t]
\begin{center}
\leavevmode
\epsfig{figure=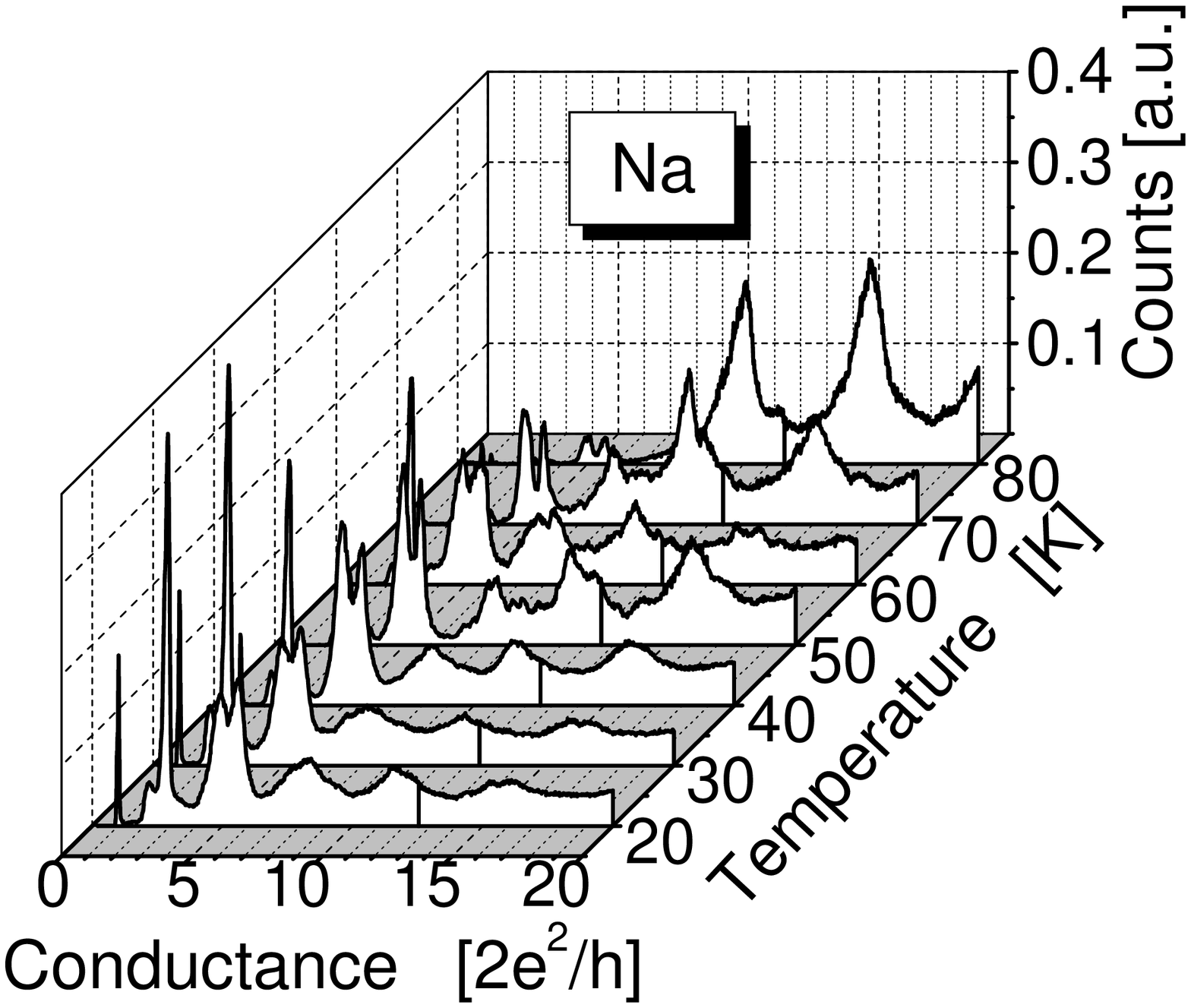,width=7.5cm}
\end{center}
\caption{Temperature evolution of sodium histograms in the
range from 0 to 20~$G_0$. The voltage bias was 10~mV and each histogram is
constructed from 1000-2000 individual scans. The amplitude has been
normalised by the total area under each histogram (a.u., arbitrary units). }
\label{fig:fig2} 
\end{figure}

Our central results are shown in Fig.~3. The histogram for sodium at temperature $T=80$~K up to $G/G_{0}=120$ is shown in Fig.~3a inset. The smooth background shown by the dotted curve was subtracted, in order to present only the oscillating part; the latter is plotted on a semi-logarithmic scale in the main panel of Fig.~3a. Up to 17 oscillations are observed, at positions which are reproduced well for $\sim 10$ different samples. The influence of the procedure of background subtraction on the peak positions is much smaller than the width of the peaks.

The radius, $R$, of the narrowest cross section of the nanowire can be obtained from the semi-classical expression for a ballistic wire \cite{Torres}, 
$$
G/G_{0}\simeq \left( {k_F R\over 2} \right)^{2}(1-{2\over k_F R}),\eqno(1)
$$
where $k_F$ is the Fermi wavevector. Using this expression we calculate the radius of the nanowire at the peak positions (in units of $k_F^{-1}$); the results are shown in Fig.~3b (open squares) for consecutively numbered peak index $i$. Thus we observe that the peak positions are periodic as a function the radius of the wire. This periodicity suggests a comparison of our experimental results with the magic numbers observed for sodium clusters. In the cluster experiments, a remarkable structure was observed in the distribution of cluster sizes produced in a supersonic expansion of sodium metal vapour 
\cite{de Heer,Knight,Martin1,Martin2,Bjornholm,Pedersen}. In the spectrum of the abundance versus the number of atoms, $N$, in the cluster, distinct maxima are observed for $N=2,8,20,40,58,92,...$. These `magic numbers' correspond to clusters having a number of valence electrons (equal to the number of atoms) which just complete a shell, where the wave functions are considered as freely propagating waves inside a spherical potential well. The magic numbers are periodically distributed as a function of the radius of the clusters, where the radius is obtained from the number of atoms, $N$, in the cluster using \cite{de Heer,Brack93} $k_F R=1.92 N^{1/3}$. In Fig.~3b  we compare the radii for the cluster magic numbers (filled circles) with the radii corresponding to the peaks in the conductance histogram (open squares). We establish a direct correspondence between preferential values for the conductance in histograms for sodium nanowires, and the cluster magic numbers.

The linear relation between the radii at cluster magic numbers and the shell number arises owing to fluctuations in the density of states as a function of $k_F R$, which in turn give rise to fluctuations in the free energy of the system, $\Omega(R)$. The periods of oscillation can be expressed in terms of semiclassical closed orbits inscribed inside the boundaries of the system \cite{Balian,Brack93}, where the path length $L$ should be an integer multiple of the Fermi wavelength $\lambda_{F}=2\pi /k_{F}$. Fluctuations in the density of states for a free electron nanowire and their influence on the free energy has been recently considered in a number of theoretical papers \cite{Staf97,Kass98,Ruiten97,Yann97,Yann98,Hopp98}. Using the semiclassical expansion of the free energy for a cylindrical wire \cite{Yann98,Hopp98} including only the three lowest order terms, we calculated the positions of the energy minima, which are plotted as a function of shell number in Fig.~3b (open triangles). The agreement with the experimental points is fairly good, and may be improved taking into account a more realistic shape of the potential well and higher-order contributions.

\begin{figure}[!b]
\begin{center}
\leavevmode
\epsfig{figure=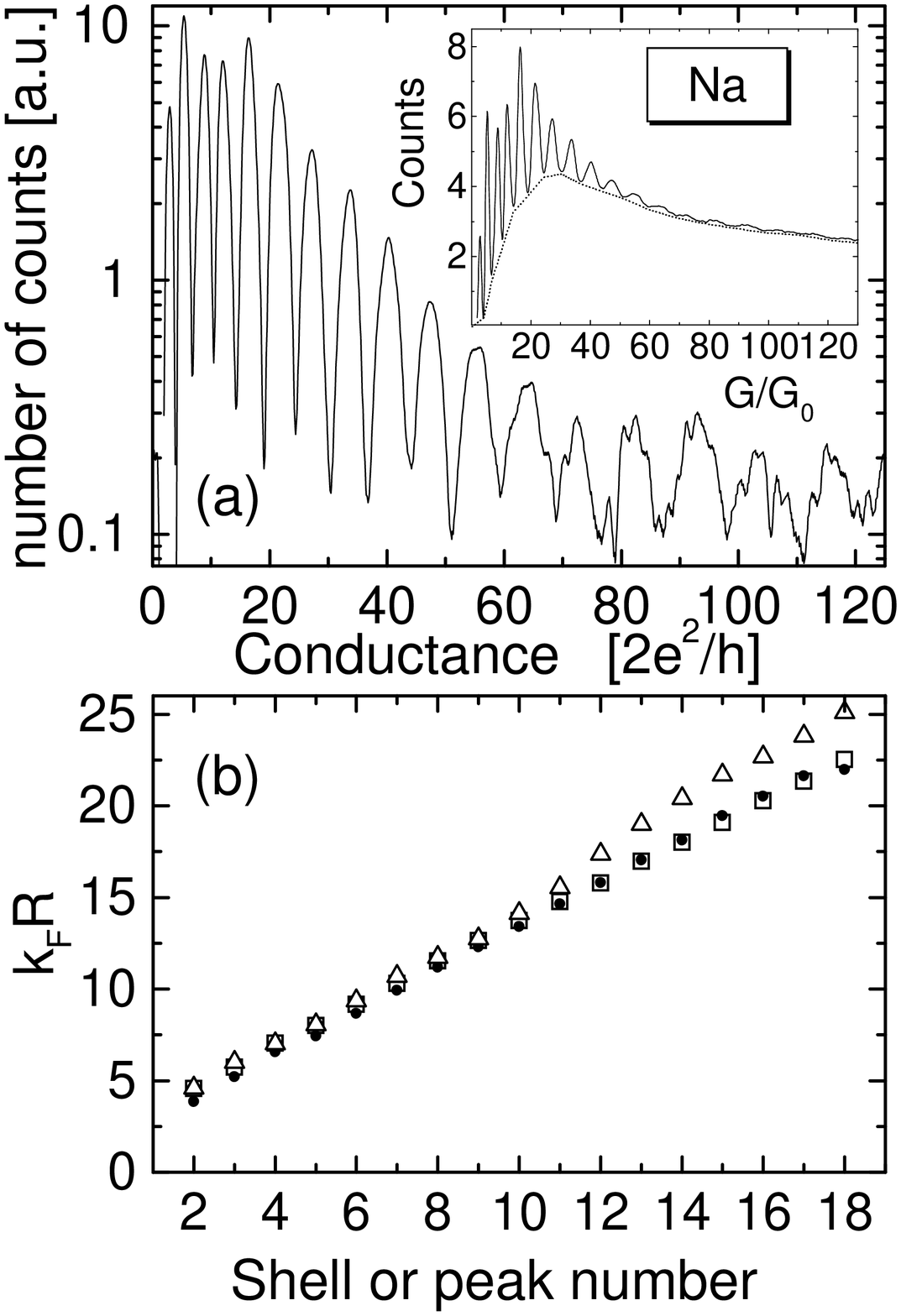,width=7.5cm}
\end{center}
\caption{ Conductance histograms for sodium, showing evidence for shell structure. {\bf (a)}, Histogram of the number of times each conductance 
is observed versus the conductance in units of $G_0=2e^2/h$. These data are for sodium at $T=80$~K and bias voltage $V=100$~mV, constructed from over 10,000 individual scans. Inset, the raw data and the smooth background (dotted curve); the background is subtracted to give the curve in the main graph. The logarithmic scale for ordinate axis helps to display the smaller-amplitude features at high conductance values. {\bf (b)}, Radius of the nanowire at the positions of the maxima in {\bf (a)} versus peak numbers (open squares), where $R$ is given in units $k_{\rm F}^{-1}$). The radii at the peak positions are compared with the radii corresponding to the magic numbers for sodium metal clusters (filled circles)\protect\cite{Martin2} and with those expected from a 
semiclassical description for the fluctuations in the free energy for the 
nanowire (open triangles)\protect\cite{Yann98,Hopp98}.}
\label{fig3}
\end{figure}

Our interpretation uses the approximately linear semiclassical relation $G(R)$ between conductance and wire cross-section, equation~(1). Corrections may arise from two mechanisms. First, back scattering on defects (or phonons) may shift the peak positions. The fact that strong scattering would tend to smear the peak structure, and the close agreement between the experimental and theoretical periodicity both suggest that this shift is small. Second, the mechanism of level-bunching leading to fluctuations $\Omega(R)$ should also give rise to fluctuation corrections to $G(R)$, which by itself would lead to peaks in the histograms even for a perfectly smooth distribution of wire diameters. Peaks due to this mechanism would be found at those points where the increase of the conductance with wire diameter is slow, and indeed we believe the peaks at low conductance, near 1, 3 and 6 $G_0$ are due to this mechanism. However, we argue that the main structure in Fig.~3 is due to the shell structure in $\Omega(R)$.

The main argument in favour of this interpretation comes from the temperature dependence shown in Fig.~2. As in the cluster experiments \cite{Martin2}, the electronic shell structure becomes observable at higher temperatures, where the increased mobility of the atoms allows the system to explore a wider range of neighbouring atomic configurations in order to find the local energy minima. At low temperatures, the nanowires are frozen into the shape, which evolves from mechanical deformations in the breaking and indenting processes. The decrease in amplitude of the sharp quantization peaks at low conductance can probably be explained by thermally induced breaking of the contact when it consists of only a few atoms. If the fluctuations in $G(R)$ were solely responsible for the observed peak structure, it could be argued that the higher temperature would favour formation of a smoother and longer wire, leading to less backscattering of electrons and less tunnelling corrections, respectively. Backscattering is responsible for the shift to lower values of the conductance peaks near 1, 3, 5 and 6~$G_0$ \cite{Ludoph}, and we find that the position of these peaks does not change with temperature. An indication of the length of the wire may be obtained from the global variation of the conductance with elongation \cite{Untiedt} and the time evolution for fixed elongation \cite{Gai}. In our experiment, both variations show that the effective wire length decreases for increasing temperature. With these arguments an interpretation of the temperature dependence of Fig.~2 in terms of $G(R)$ fluctuations alone is ruled out.

The correspondence between the shell structure in clusters and in nanowires can be explained by comparing the energy levels for a three-dimensional spherical potential well and a two-dimensional cylindrical geometry. The levels are obtained from the zeros of the spherical and ordinary Bessel functions, respectively. Apart from a small constant shift, the distribution of levels for the two systems has a very similar structure, with gaps and bunches of levels occurring at the same positions. This explains the striking similarity of the magic radii for the clusters and nanowires observed in Fig.~3b.

\noindent\textbf{Acknowledgements}. We thank L.J. de Jongh for his continuous support.

\bigskip
\noindent
Correspondence should be addressed to J.M.v.R. (e-mail:
Ruitenbe@Phys.LeidenUniv.nl).


\begin{references}

\bibitem{Balian}  Balian, R. \& Bloch, C. Distribution of eigenfrequencies
for the wave equation in a finite domain: III. Eigenfrequency density
oscillations. {\it Ann. Phys.(N.Y.)} {\bf 69}, 76-7-160 (1972).

\bibitem{Bohr}  Bohr, \AA . \& Mottelson, B.R. {\it Nuclear Structure},
Vol.II (Benjamin, Reading, MA, 1975).

\bibitem{de Heer}  de Heer, W.A. The physics of simple metal clusters:
experimental aspects and simple models. {\it Rev. Mod. Phys.} {\bf 65},
611--676 (1993).

\bibitem{Brack93}  Brack, M. The physics of simple metal clusters:
self-consistent jellium model and semiclassical approaches.{\it \ Rev. Mod.
Phys.} {\bf 65}, 677--732 (1993).

\bibitem{Agrait93}  Agra\"\i t, N., Rodrigo, J.G. \& Vieira, S. Conductance
steps and quantization in atomic-size contacts. {\it Phys. Rev. B} {\bf 47},
12345--12348 (1993).

\bibitem{Muller92}  Muller, C.J., van Ruitenbeek, J.M., \& de Jongh, L.J.
Experimental observation of the transition from weak link to tunnel
junction. {\it Physica C} {\bf 191}, 485--504 (1992).

\bibitem{Krans95}  Krans, J.M. {\it et al.}\
The signature of conductance quantization in metallic
point contacts. {\it Nature }{\bf 375}, 767--769 (1995).

\bibitem{Torres}  Torres, J.A., Pascual, J.I. \& S\'{a}enz, J.J. 
Theory of conduction through narrow constrictions in a three-dimensional 
electron gas, {\it Phys. Rev. B} {\bf 49}, 16581--16584 (1994).

\bibitem{Knight} Knight, W.D. {\it et al.}\
Electronic shell structure and abundances of sodium clusters, {\it Phys.\ Rev.\ Lett.\ }\ {\bf 52}, 2141--2143 (1984).

\bibitem{Martin1} Martin, T.P. {\it et al.}\
Observation of electronic shells and shells of atoms in large Na clusters, {\it Chem.\ Phys.\ Lett.}\ {\bf 172}, 209--213 (1990).

\bibitem{Martin2} Martin, T.P. {\it et al.}\
Electronic shell structure of laser-warmed Na clusters, {\it Chem.\ Phys.\ Lett.}\ {\bf 186}, 53--57 (1991).

\bibitem{Bjornholm} Bj{\o}rnholm, S. {\it et al.}\
Mean-field quantization of several hundred electrons in sodium metal clusters, {\it Phys.\ Rev.\ Lett.\ }\ {\bf 65}, 1627--1630 (1990). 

\bibitem{Pedersen} Pedersen, J. {\it et al.}\
Observation of quantum supershells in clusters of sodium atoms, {\it Nature }\ {\bf 353}, 733--735 (1991).

\bibitem{Staf97}  Stafford, C.A., Baeriswyl, D. \& B\"{u}rki, J. Jellium
model of metallic nanocohesion. {\it Phys. Rev. Lett.} {\bf 79}, 2863--2866
(1997).

\bibitem{Kass98}  Kassubek, F., Stafford, C.A. \& Grabert, H. Force, charge,
and conductance of an ideal metallic nanowire, {\it Phys. Rev. B} {\bf 59}, 7560--7574 (1999).

\bibitem{Ruiten97}  van Ruitenbeek, J.M., Devoret, M.H., Esteve, D. \&
Urbina, C. Conductance quantization in metals: The influence of subband
formation on the relative stability of specific contact diameters. {\it %
Phys. Rev. B }{\bf 56}, 12566--12572 (1997).

\bibitem{Yann97}  Yannouleas, C. \& Landman, U., On mesoscopic forces and
quantized conductance in model metallic nanowires. {\it J. Phys. Chem. B }%
{\bf 101}, 5780--5783 (1997).

\bibitem{Yann98}  Yannouleas, C., Bogachek, E.N. \& Landman, U. Energetics,
forces, and quantized conductance in jellium-modeled metallic nanowires. 
{\it Phys. Rev. B }{\bf 57}, 4872--4882 (1998).

\bibitem{Hopp98}  H\"{o}ppler, C. \& Zwerger, W. Quantum fluctuations in the cohesive force of metallic nanowires. {\it Phys. Rev. B }{\bf 59}, R7849--R7851 (1999).

\bibitem{Ludoph} Ludoph, B. {\it et al.}\ Evidence for saturation of channel transmission from conductance fluctuations in atomic-size point contacts. {\it Phys.\ Rev.\ Lett.\ } {\bf 82}, 1530--1533 (1999). 

\bibitem{Untiedt} Untiedt, C. {\it et al.}\
Fabrication and characterization of metallic nanowires
{\it Phys.\ Rev.\ B} {\bf 56}, 2154--2160 (1997).

\bibitem{Gai} Gai, Z., {\it et al.}\
Spontaneous breaking of nanowires between a STM tip and the Pb(110) surface.
{\it Phys.\ Rev.\ B} {\bf 58}, 2185--2190 (1998).


\end{references}
\end{document}